\documentclass[aps,prl,twocolumn,groupedaddress]{revtex4-1}
\usepackage{graphicx}
\begin{document}

% Use the \preprint command to place your local institutional report
% number in the upper righthand corner of the title page in preprint mode.
% Multiple \preprint commands are allowed.
% Use the 'preprintnumbers' class option to override journal defaults
% to display numbers if necessary
%\preprint{}

%Title of paper
\title{Andreev Reflection Like Enhancement Above Bulk $T_c$ in Electron Underdoped Iron Arsenides}

% repeat the \author .. \affiliation  etc. as needed
% \email, \thanks, \homepage, \altaffiliation all apply to the current
% author. Explanatory text should go in the []'s, actual e-mail
% address or url should go in the {}'s for \email and \homepage.
% Please use the appropriate macro foreach each type of information

% \affiliation command applies to all authors since the last
% \affiliation command. The \affiliation command should follow the
% other information
% \affiliation can be followed by \email, \homepage, \thanks as well.
\author{H. Z. Arham$^1$}
\email[E-mail:]{arham1@illinois.edu}
\author{C. R. Hunt,$^1$ J. Gillett,$^2$ S. D. Das,$^2$ S. E. Sebastian$^2$}
\author{D. Y. Chung,$^3$ M. G. Kanatzidis$^3$}
\author{L. H. Greene$^1$}

%\homepage[]{Your web page}
%\thanks{}
%\altaffiliation{}
\affiliation{\\\textsuperscript{$1$}Department of Physics and the Frederick Seitz Material Research Laboratory, University of Illinois at Urbana-Champaign, Urbana, Illinois 61801, USA
\\\textsuperscript{$2$}Cavendish Laboratory, J. J. Thomson Ave, University of Cambridge, UK
\\\textsuperscript{$3$}Materials Science Division, Argonne National Laboratory, Argonne, IL 60439, USA}

%Collaboration name if desired (requires use of superscriptaddress
%option in \documentclass). \noaffiliation is required (may also be
%used with the \author command).
%\collaboration can be followed by \email, \homepage, \thanks as well.
%\collaboration{}
%\noaffiliation

\date{\today}

\begin{abstract}
We use point contact spectroscopy (PCS) to probe the superconducting properties of electron doped $\rm{Ba(Fe_{1-x}Co_x)_2As_2}$ ($\rm{x = 0.05, 0.055, 0.07, 0.08}$) and hole doped $\rm{Ba_{0.8}K_{0.2}Fe_2As_2}$. PCS directly probes the low energy density of states via Andreev reflection, revealing two distinct superconducting gaps in both compound families. Apart from the electron underdoped $\rm{Ba(Fe_{1-x}Co_{x})_2As_2}$, the excess current due to Andreev reflection for the compounds follows the typical BCS temperature dependence. For underdoped $\rm{Ba(Fe_{1-x}Co_{x})_2As_2}$, the temperature dependence of the excess current deviates from that of BCS, developing a tail at higher temperatures and surviving above bulk $T_c$. Possible explanations for this anomalous behavior are explored. 
\end{abstract}

% insert suggested PACS numbers in braces on next line
\pacs{}
% insert suggested keywords - APS authors don't need to do this
%\keywords{}

%\maketitle must follow title, authors, abstract, \pacs, and \keywords
\maketitle

% body of paper here - Use proper section commands
% References should be done using the \cite, \ref, and \label commands

Point contact spectroscopy (PCS) \cite{Naidyuk} proves to be an extremely useful spectroscopic technique for studying conventional and unconventional superconductors since it is sensitive to the magnitude and symmetry of the superconducting order parameter. A point contact junction consists of a nanoscale metallic contact with a superconductor, with transport across the junction dominated by Andreev reflection \cite{Andreev}. The density of states may be directly extracted from the conductivity using the Blonder-Tinkham-Klapwijk (BTK) model \cite{Blonder}. PCS was instrumental in determining the precise location of the line nodes for the heavy fermion compound $\rm{CeCoIn_5}$ \cite{WKPark}, and in providing direct evidence for the multi-gap nature of the superconductor $\rm{MgB_2}$ \cite{MgB2}. 

A number of research groups have utilized PCS to study the iron based superconductors. The results are well summarized in a recent review article by Daghero et al \cite{ROPP}. For the $\rm{Ba(Fe_{1-x}Co_x)_2As_2}$ and $\rm{Ba_{1-x}K_{x}Fe_2As_2}$ families, measurements on near optimal and overdoped samples have revealed the existence of multiple gaps consistent with s-wave pairing \cite{Samuely, Tortello}. To our knowledge, no results have been reported for underdoped compounds, which constitute the most fascinating region of the 122 family phase diagram. Underdoped compounds exhibit a coexistence of magnetism and superconductivity at low temperatures \cite{Johnston} as well as nematic fluctuations in their normal state \cite{Fisher}. 

In this paper we present Andreev reflection data indicating multiple s-wave superconducting gaps for electron underdoped $\rm{Ba(Fe_{1-x}Co_x)_2As_2}$ ($\rm{x = 0.05, 0.055}$) and hole underdoped $\rm{Ba_{0.8}K_{0.2}Fe_2As_2}$. We also present data on near optimal electron doped $\rm{Ba(Fe_{1-x}Co_x)_2As_2}$ ($\rm{x = 0.07, 0.08}$) that is in agreement with the published PCS literature. 

We fit our lowest temperature data using the extended BTK model with two s-wave superconducting gaps \cite{Brinkman}. All the point contact junctions show split Andreev peaks and none of the compounds have a maximum at zero bias voltage. This indicates that the superconducting order parameter does not have any nodes and the Fermi surfaces are fully gapped. 

We define the superconducting transition by two temperatures: $T_c^{onset}$, for when the resistive transition starts, and $T_c^{bulk}$, for when it falls by 90$\%$ of the value at $T_c^{onset}$. The electron underdoped compounds show an Andreev reflection-like conductance enhancement between $T_c^{bulk}$ and $T_c^{onset}$ which we argue most likely arises from inhomogenous doping effects. For the rest of the compounds, the temperature evolution of the excess current due to Andreev reflection appears to follow the standard BCS like behavior and disappears at $T_c^{bulk}$.

\begin{figure}[thbp]
		\includegraphics[scale=0.56]{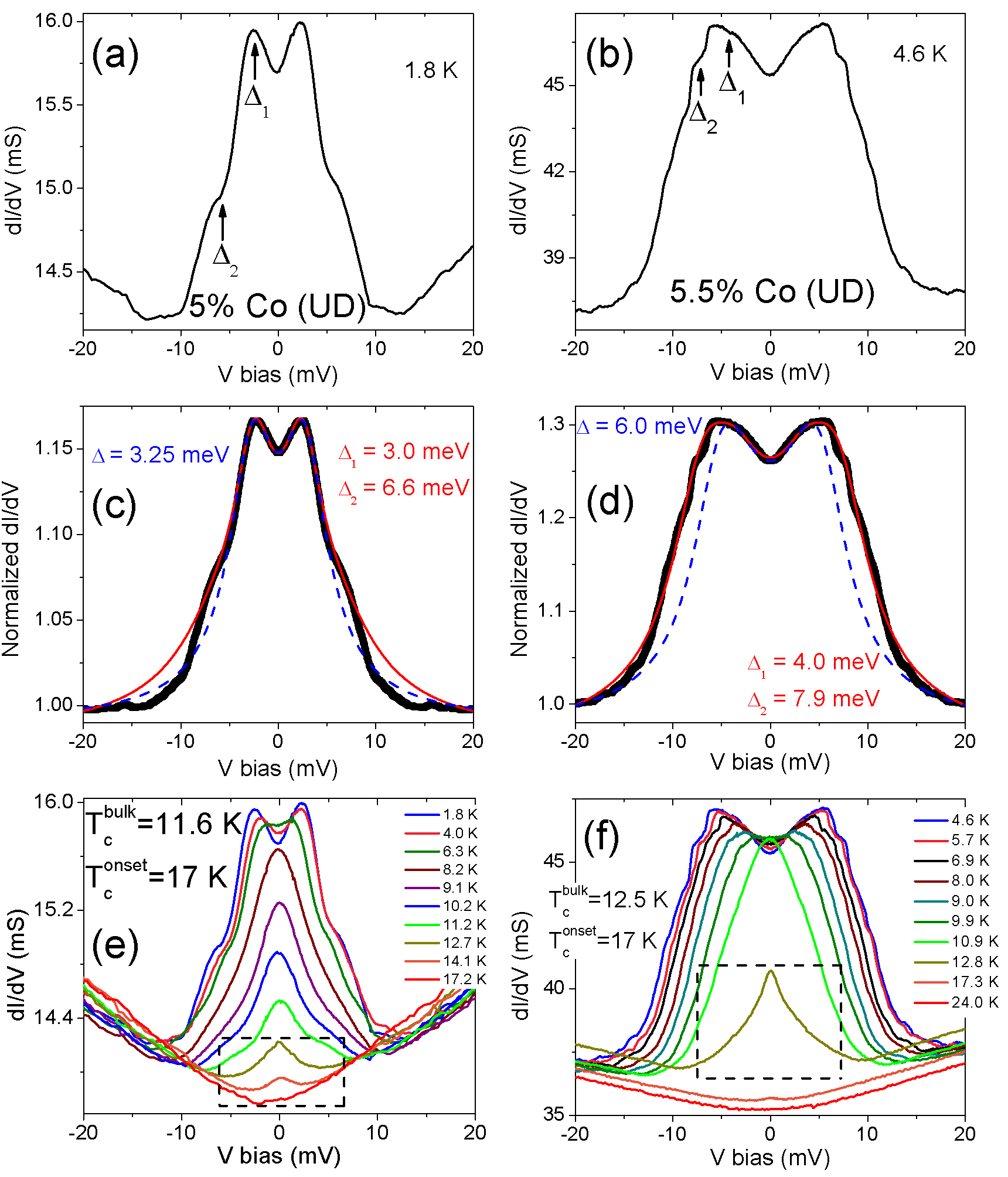}
	\caption{(color online) $dI/dV$ for $\rm{Ba(Fe_{0.95}Co_{0.05})_2As_2}$ (left column) and $\rm{Ba(Fe_{0.945}Co_{0.055})_2As_2}$ (right column). (a, b) Low temperature Andreev reflection spectra, the arrows point out the features corresponding to the two gaps. (c, d) The data shown in (a, b) have been normalized and fit to one band and two band BTK models. The one band fit (dashed blue line) fails to account for the larger gap. The two band fit (solid red line) is able to track the experimental data better. The gap values obtained for the 5$\%$ Co sample are $\rm{\Delta_1 = 3.0 meV}$ and $\rm{\Delta_2 = 6.6 meV}$ while those for the 5.5$\%$ Co sample are $\rm{\Delta_1 = 4.0 meV}$ and $\rm{\Delta_2 = 7.9 meV}$. (e, f) Temperature evolution of the Andreev reflection spectra.}
	\label{fig:100}
\end{figure}  

Single crystals of $\rm{Ba(Fe_{1-x}Co_x)_2As_2}$ are grown out of FeAs flux \cite{Sebastian, Gillett} (x = 0.05, 0.055, 0.07, 0.08). $\rm{Ba_{0.8}K_{0.2}Fe_2As_2}$ crystals are grown in Sn flux \cite{Duck}. Metallic junctions are formed on freshly cleaved c-axis crystal surfaces and $dI/dV$ across each junction is measured using a standard four-probe lock-in technique. Junctions are constructed via the soft PCS method \cite{Arham}.  

Fig. 1 presents $dI/dV$ curves for $\rm{Ba(Fe_{0.95}Co_{0.05})_2As_2}$ (left column, $T_c^{bulk}$ = 11.6 K, $T_c^{onset}$ = 17 K) and $\rm{Ba(Fe_{0.945}Co_{0.055})_2As_2}$ (right column, $T_c^{bulk}$ = 12.5 K, $T_c^{onset}$ = 17 K). Fig. 1a and 1b show the $dI/dV$ raw data at the lowest temperatures attained. The Andreev spectra clearly points to the presence of two superconducting gaps, as indicated with arrows. Fig. 1c and 1d show BTK fits to the normalized data from Fig. 1a and 1b, respectively. The dotted blue line is an isotropic s-wave single band BTK fit. While it provides a good fit to the features corresponding to the smaller gap, it cannot account for the features corresponding to the larger gap. A two band BTK approach, solid red line, is required to accurately describe the experimental data. The parameters in the fit are the two superconducting gaps $\Delta_1$ and $\Delta_2$, the Dynes lifetime broadening parameter for these gaps $\Gamma_1$ and $\Gamma_2$ \cite{Dynes}, the transparency of the junction for each gap $Z_1$ and $Z_2$, and the weight of the first gap $w$. (The weight of the second gap becomes $1-w$). For the best fits, $Z_1$ and $Z_2$ are close to each other but not identical. This might be due to the different Fermi velocities for the different Fermi surfaces resulting in unequal effective barrier strengths. The ratio $\Gamma/\Delta$ for both gaps are also similar.

The parameters for all our fits are given in Table 1. For $\rm{Ba(Fe_{0.95}Co_{0.05})_2As_2}$ $\rm{\Delta_1 = 3.0}$ meV and $\rm{\Delta_2 = 6.6}$ meV while for $\rm{Ba(Fe_{0.945}Co_{0.055})_2As_2}$ $\rm{\Delta_1 = 4.0}$ meV and $\rm{\Delta_2 = 7.9}$ meV. Fig. 1e and 1f show the raw $dI/dV$ temperature evolution curves of the Andreev spectra for these two junctions. The dashed black rectangle is highlighting the conductance enhancement that is detected between $T_c^{bulk}$ and $T_c^{onset}$. 

\begin{figure}[thbp]
		\includegraphics[scale=0.56]{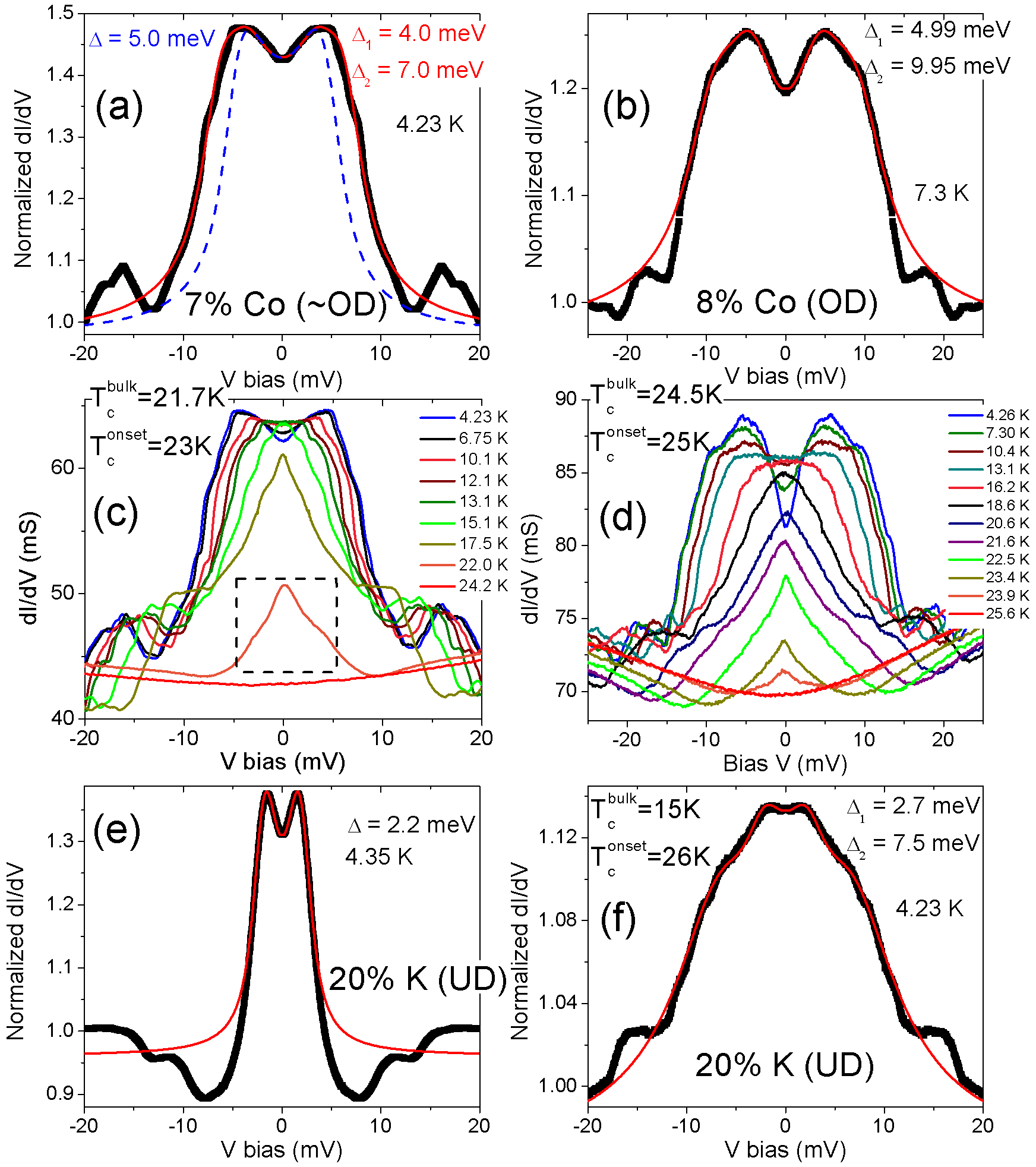}
	\caption{(color online) (a) Low temperature Andreev reflection spectra for $\rm{Ba(Fe_{0.93}Co_{0.07})_2As_2}$ has been normalized and fit to one band and two band BTK models. The two band BTK model provides a much better fit with $\rm{\Delta_1 = 4.0 meV}$ and $\rm{\Delta_2 = 7.0 meV}$. (b) Two band BTK fit for $\rm{Ba(Fe_{0.92}Co_{0.08})_2As_2}$ provides $\rm{\Delta_1 = 4.99 meV}$ and $\rm{\Delta_2 = 9.95 meV}$.(c, d) Temperature evolution of the Andreev reflection spectra for the junction in (a, b). (e) Single band BTK fit for $\rm{Ba_{0.8}K_{0.2}Fe_2As_2}$ ($\rm{\Delta = 2.2 meV}$). (f) Two band BTK fit for a different junction on $\rm{Ba_{0.8}K_{0.2}Fe_2As_2}$. The gap values are $\rm{\Delta_1 = 2.7 meV}$, $\rm{\Delta_2 = 7.5 meV}$.}
	\label{fig:100}
\end{figure}  

\begin{table*}
 \caption{\label{Label}}
 \begin{ruledtabular}
 \begin{tabular}{c c c c c c c c c c c c}
 Crystal & $\Delta_1$ & $\Delta_2$ & $Z_1$ & $Z_2$ & $\Gamma_1/\Delta_1$ & $\Gamma_2/\Delta_2$ & w & $T_c^{onset}$ & $T_c^{bulk}$ & $T_S$ & $T_N$\\
 \hline
 $\rm{Ba(Fe_{0.95}Co_{0.05})_2As_2}$ (e UD)    & 3.0 meV & 6.6 meV & 0.395 & 0.40 & 0.56 & 0.62 & 0.67 & 17 K & 11.6 K & 78 K & 70 K \\
 $\rm{Ba(Fe_{0.945}Co_{0.055})_2As_2}$ (e UD)  & 4.0 meV & 7.9 meV & 0.31  & 0.35 & 0.25 & 0.30 & 0.10 & 17 K & 12.5 K & 75 K & 63 K\\
 $\rm{Ba(Fe_{0.93}Co_{0.07})_2As_2}$ (e OD)    & 4.0 meV & 7.0 meV & 0.28  & 0.30 & 0.20 & 0.15 & 0.20 & 23 K & 21.7 K & - & -\\
 $\rm{Ba(Fe_{0.92}Co_{0.08})_2As_2}$ (e OD)    & 4.99 meV & 9.95 meV & 0.39 & 0.435  & 0.36 & 0.23 & 0.50 & 25 K & 24.5 K & - & -\\
 $\rm{Ba_{0.8}K_{0.2}Fe_2As_2}$ (h UD)         & 2.2 meV & - & 0.373 & -  & 0.18 & - & 1 & 26 K & 15 K & 90 K & 90 K\\
 $\rm{Ba_{0.8}K_{0.2}Fe_2As_2}$ (h UD)        & 2.7 meV & 7.5 meV & 0.32 & 0.45  & 0.57 & 0.53 & 0.37 & 26 K & 15 K & 90 K & 90 K\\
 \end{tabular}
 \end{ruledtabular}
 \end{table*}

Fig. 2a and c show $dI/dV$ data and BTK fits for $\rm{Ba(Fe_{0.93}Co_{0.07})_2As_2}$ ($T_c^{bulk}$ = 21.7 K, $T_c^{onset}$ = 23 K).  The two band BTK model (solid red line $\rm{\Delta_1 = 4.0}$ meV, $\rm{\Delta_2 = 7.0}$ meV) provides a closer fit to the experimental data shown as opposed to the one band BTK model (dotted blue line). Note here also a conductance enhancement just above $T_c^{bulk}$. 

For $\rm{Ba(Fe_{0.92}Co_{0.08})_2As_2}$ ($T_c^{bulk}$ = 24.5 K, $T_c^{onset}$ = 25 K) the two band BTK fit (solid red line Fig. 2b) gives $\rm{\Delta_1 = 4.99}$ meV, $\rm{\Delta_2 = 9.95}$ meV. Fig. 2d shows the temperature evolution of the Andreev spectra of this junction. 

For near optimal doped $\rm{Ba(Fe_{1-x}Co_x)_2As_2}$ our gap values are in good agreement with those reported in the literature for PCS \cite{Tortello, ROPP}, scanning tunneling microscopy \cite{Teague}, and angle resolved photoemission spectroscopy \cite{Terashima}.    
  
Fig. 2e and f show $dI/dV$ for two different junctions on $\rm{Ba_{0.8}K_{0.2}Fe_2As_2}$. The superconducting transition is very broad, with $T_c^{onset}$ = 26 K and $T_c^{bulk}$ = 15 K. Fig. 2e shows the data can be fit with a single superconducting gap ($\rm{\Delta = 2.2}$ meV), while Fig. 2f shows clear features corresponding to two superconducting gaps ($\rm{\Delta_1 = 2.7}$ meV, $\rm{\Delta_2 = 7.5}$ meV).

The Fermi surfaces of these compounds are quasi two dimensional sheets with elliptical electron pockets centered at $(0,\pi)$ and $(\pi,0)$ and near circular hole pockets at the $\Gamma$ point \cite{Chubukov}. For $\rm{Ba_{0.6}K_{0.4}Fe_2As_2}$, a small energy gap is observed on hole pocket $\beta$ while nearly equal large energy gaps are found on hole pocket $\alpha$ and electron pocket $\gamma$ \cite{Ding}. However, the Fermi surface of $\beta$ is 4$-$6 times larger than that of $\alpha$ and $\gamma$. It is plausible that on occasion our point contacts pick up the gap only from $\beta$ causing our spectra to be a good fit to the single gap BTK model.

All the fits we have shown in the paper assume isotropic s-wave superconducting gaps. We have also not included any coupling between the two bands in the multi-gap fits. Extensions to the BTK theory have been proposed to incorporate interference and phase difference between the two superconducting bands \cite{Golubov, Sudbo}. A single band BTK fit has three free parameters ($\Delta$, $Z$, $\Gamma$) while an independent two band BTK fit has seven ($\Delta_1$, $Z_1$, $\Gamma_1$, $\Delta_2$, $Z_2$, $\Gamma_2$, w). The independent two band BTK model is giving quite good fits to the experimental data, albeit the values of the parameters are somewhat degenerate, the gap sizes can be influenced within $\pm$ 0.5 mV by changing the relative weight of the bands. We have found that adding interference and a phase difference between the bands adds two more free parameters and increase uncertainty in the extracted results without improving fit quality. In the transparent junction (low Z limit) data presented here, the $dI/dV$ spectra predicted by the independent and interfering band models do not differ greatly. The interfering band models would be useful to differentiate between $s_{++}$ and $s_{+-}$ symmetry if the barrier strength Z for the same junction could be varied systematically from the metallic to the tunneling regime. However, this is very hard to achieve experimentally. 

\begin{figure}[thbp]
		\includegraphics[scale=0.56]{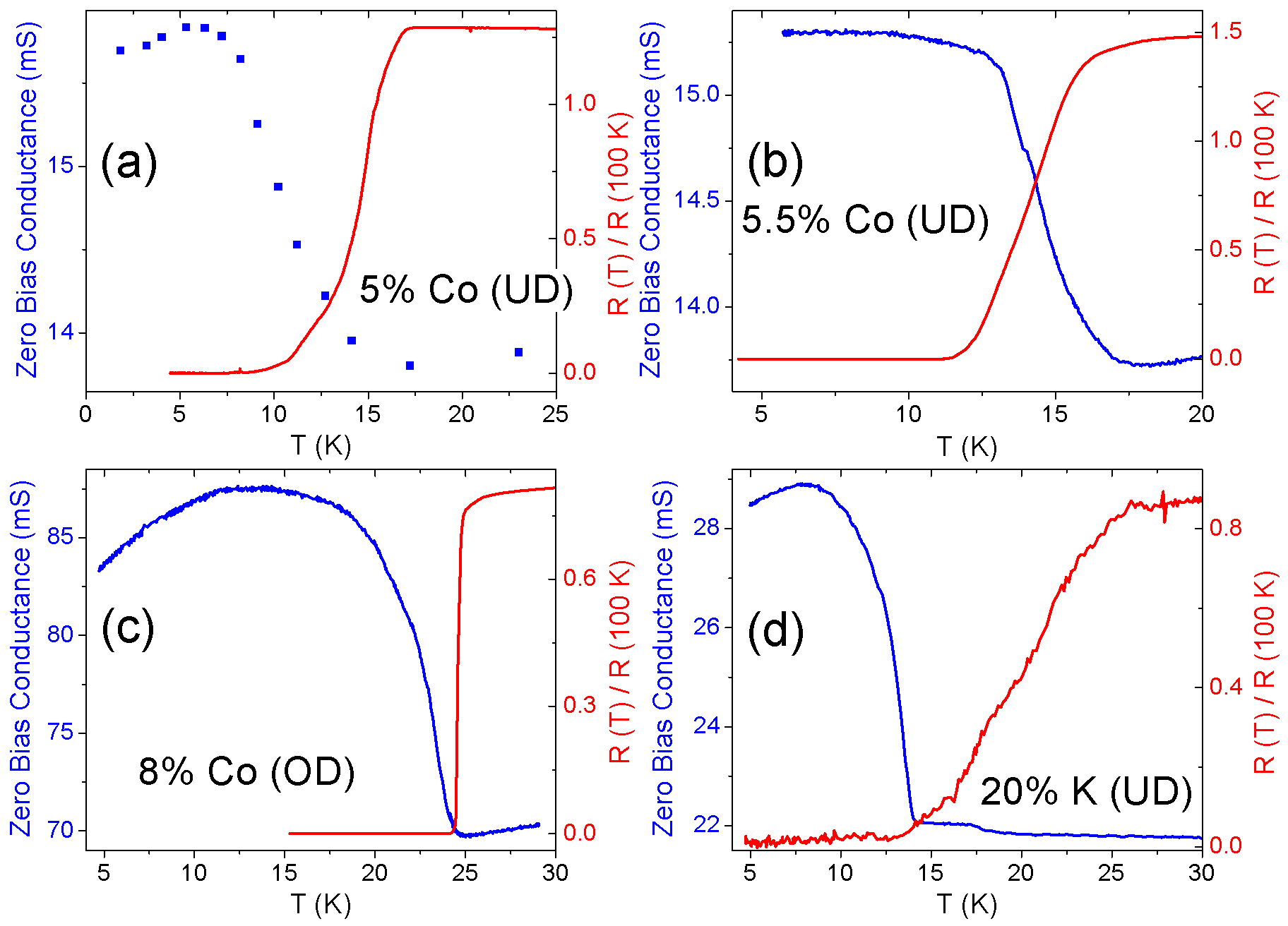}
	\caption{(color online) Comparing the zero bias conductance (blue) and bulk resistivity (red) curves. (a, b) For the electron underdoped compound $\rm{Ba(Fe_{0.945}Co_{0.055})_2As_2}$ and $\rm{Ba(Fe_{0.95}Co_{0.05})_2As_2}$, the superconducting transitions are wide and the zero bias conductance starts to rise close to $T_c^{onset}$. (c) For the near optimal doped compound $\rm{Ba(Fe_{0.92}Co_{0.08})_2As_2}$ the superconducting transition is narrow. (d) For $\rm{Ba_{0.8}K_{0.2}Fe_2As_2}$, like the electron underdoped compounds the transition is broad but the zero bias conductance only starts to rise close to $T_c^{bulk}$.} 
	\label{fig:100}
\end{figure}  

In Fig. 3 we plot the zero bias conductance and bulk resistivity on the same temperature scale for some of our junctions. For underdoped $\rm{Ba(Fe_{1-x}Co_x)_2As_2}$, (Fig. 3a, b), the superconducting transitions are broad and the zero bias conductances of the point contacts start rising near $T_c^{onset}$. For the near optimal doped sample, (Fig. 3c), the superconducting transition is narrow. For underdoped $\rm{Ba_{0.8}K_{0.2}Fe_2As_2}$, (Fig. 3d), the transition is again broad. However, in this case, the zero bias conductance starts to rise closer to $T_c^{bulk}$ as opposed to $T_c^{onset}$. 

\begin{figure}[thbp]
		\includegraphics[scale=0.8]{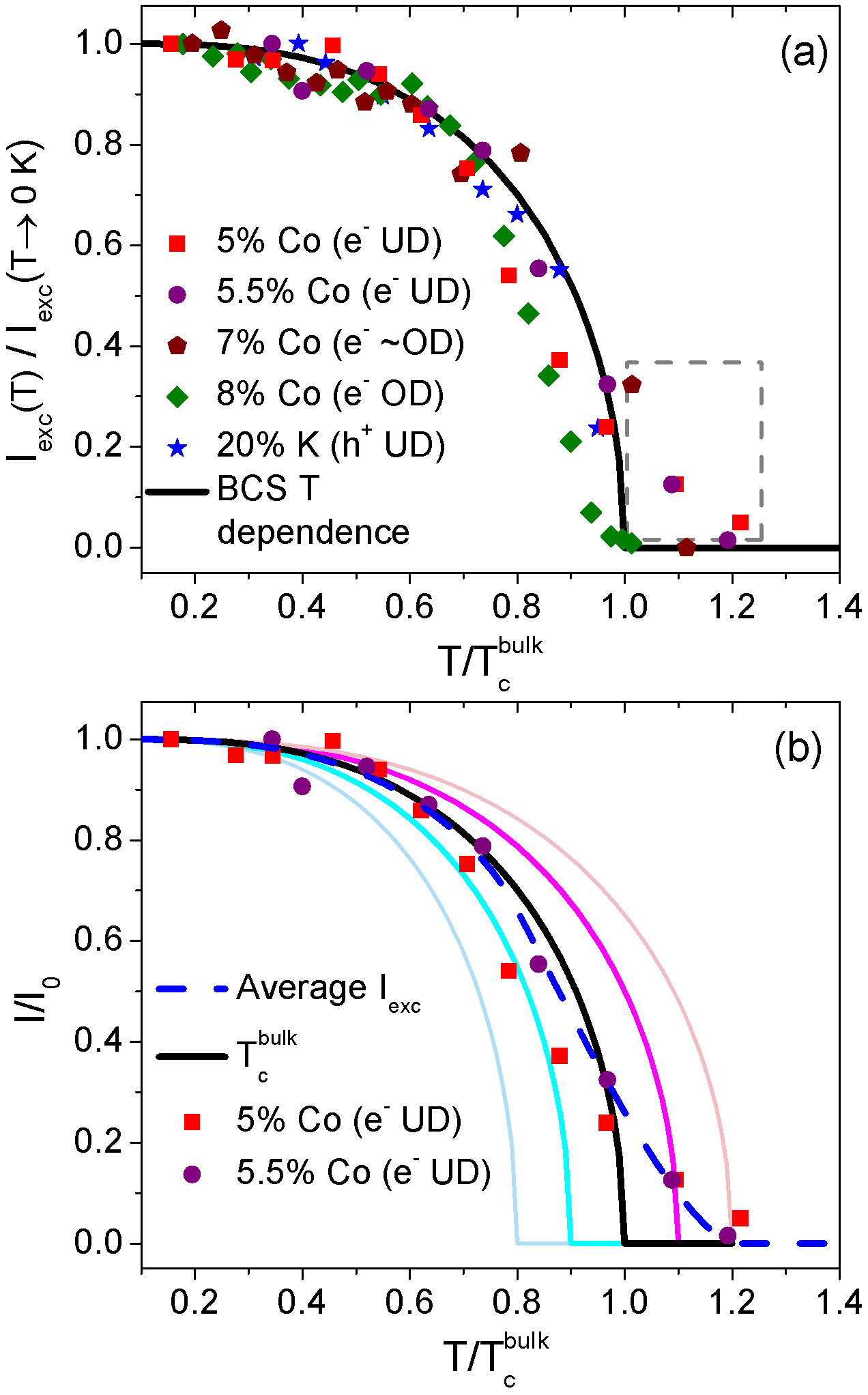}
	\caption{(color online) (a) The temperature evolution of the excess current, $I_{exc}$, for all our samples. The solid black line shows the dependence for weakly coupled s-wave BCS superconductors. The temperature has been normalized to bulk $T_c$. Apart from the electron underdoped compounds, reasonable fits are obtained. For them, $I_{exc}$ initially follows the BCS trend before developing a tail at higher temperatures. The dashed gray rectangle is highlighting this anomaly. (b) The dashed blue line is $I_{exc}$ calculated by assuming that the junction is comprised of multiple point contacts in parallel and microscopic inhomogeneities in the Co doping give rise to a Gaussian distribution function centered at $T_c^{bulk}$ for the local $T_c$ of the point contacts.}
	\label{fig:100}
\end{figure}  

The approximate temperature dependence of the energy gap for weakly coupled s-wave BCS superconductors may be given by $\rm{\Delta=\Delta_0tanh[1.74* \sqrt{T_c/T-1}]}$. As the temperature is increased, the Andreev reflection signal decreases with a concomitant increase in the thermal broadening in the $dI/dV$ curves. The gap values extracted by BTK fits develop larger and larger error bars and the smaller gap becomes especially hard to distinguish. Instead of plotting the temperature evolution of the extracted $\Delta$ values, we therefore focus on the excess current, $I_{exc}$, due to Andreev reflection. From the BTK theory \cite{Blonder} for s-wave superconductors, $I_{exc}$ has the same temperature dependence as $\Delta$ ($I_{exc}\propto\Delta/R_{junction}$). We calculate $I_{exc}$ by integrating the normalized $dI/dV$ curves over $\mathrm{\pm(V>>\Delta)}$ \cite{Supplement}. Fig. 4a shows $I_{exc}$ vs. $T$. To compare each doping, we normalize $T_c$ and low temperature $I_{exc}$ to 1. The near optimally doped $\rm{Ba(Fe_{0.93}Co_{0.07})_2As_2}$, $\rm{Ba(Fe_{0.92}Co_{0.08})_2As_2}$ and hole underdoped $\rm{Ba_{0.8}K_{0.2}Fe_2As_2}$ crystals show a reasonable agreement with a BCS temperature dependence. 

Analysis of the data taken on the electron underdoped $\rm{Ba(Fe_{0.95}Co_{0.05})_2As_2}$ and $\rm{Ba(Fe_{0.945}Co_{0.055})_2As_2}$ crystals is more complex. While Andreev spectra on the optimally and hole underdoped crystals exhibit $I_{exc}$ close to $T_c^{bulk}$, the data on electron underdoped compounds (5$\%$ and 5.5$\%$ Co doping) exhibit $I_{exc}$ at $T_c^{onset}$. The superconducting transition for these crystals is broad; 4.5-5 K. In Fig. 4a the solid black line is the BCS behavior vs. temperature normalized to $T_c^{bulk}$. Data from $\rm{Ba(Fe_{0.945}Co_{0.055})_2As_2}$ (purple circles) follows the fit up to $T_c^{bulk}$ after which $I_{exc}$ remains enhanced up to $T_c^{onset}$. $\rm{Ba(Fe_{0.95}Co_{0.05})_2As_2}$ (red squares) follows a similar trend.  

Microscopic variations in the Co doping may be used to explain why the electron underdoped $\rm{Ba(Fe_{1-x}Co_{x})_2As_2}$ crystals show $I_{exc}$ above their bulk $T_c$. Our soft PCS junctions are comprised of multiple point contacts and the conductivity from each adds to give the measured $I_{exc}$. We assume a Gaussian distribution function for the local $T_c$ of the point contacts centered at $T_c^{bulk}$ \cite{Supplement} and calculate the resulting $I_{exc}$ (Figure 4b). The simulated curve (dashed blue line) reproduces the experimentally observed $I_{exc}$ above $T_c^{bulk}$ quite well. The solid lines are those calcualated for multiple point contacts (with different $T_c$ values) whose weighted sum gives the total $I_{exc}$.   

An alternate explanation is that this enhancement above bulk $T_c$ is due to a novel scattering mechanism. Such scenarios have previously been reported in $\rm{FeTe_{0.55}Se_{0.45}}$ (spin fluctuations) \cite{FeTeSe} and $\rm{Ba(Fe_{1-x}Co_{x})_2As_2}$ (phase-incoherent superconducting pairs) \cite{Goutam}. Data on underdoped $\rm{Ba(Fe_{1-x}Co_{x})_2As_2}$ crystals show orbital fluctuations in their normal state, while those on optimally doped $\rm{Ba(Fe_{1-x}Co_{x})_2As_2}$ and $\rm{Ba_{0.8}K_{0.2}Fe_2As_2}$ crystals do not \cite{Arham, Weicheng}. Compounds exhibiting orbital fluctuations above the structural phase transition show $I_{exc}$ above $T_c^{bulk}$.  

To conclude, we have presented Andreev reflection PCS $dI/dV$ data for $\rm{Ba(Fe_{1-x}Co_{x})_2As_2}$ ($\rm{x = 0.05, 0.055, 0.07, 0.08}$) and  $\rm{Ba_{0.8}K_{0.2}Fe_2As_2}$. All junctions are made along the c-axis. Our spectra provide clear evidence for multiple, nodeless, s-wave superconducting gaps. The values of the two gaps may be extracted by using the independent multiband BTK model. Apart from underdoped $\rm{Ba(Fe_{1-x}Co_{x})_2As_2}$, the temperature evolution of the excess current for the crystals is well described by the BCS temperature dependence. The excess current for underdoped $\rm{Ba(Fe_{1-x}Co_{x})_2As_2}$ exhibits excess conductance at higher temperatures that survives above the bulk $T_c$. The shape of $I_{exc}$ vs. T can be simulated assuming microscopic inhomogeneity in the Co doping in the crystals.

We acknowledge W. K. Park for useful discussions. This work is supported as part of the Center for Emergent Superconductivity, an Energy Frontier Research Center funded by the US Department of Energy, Office of Science, Office of Basic Energy Sciences under Award No. DE-AC0298CH1088. University of Cambridge is supported by EPSRC, Trinity College, the Royal Society and the Commonwealth Trust.

\bibliography{myrefs}

%merlin.mbs apsrev4-1.bst 2010-07-25 4.21a (PWD, AO, DPC) hacked
%Control: key (0)
%Control: author (8) initials jnrlst
%Control: editor formatted (1) identically to author
%Control: production of article title (-1) disabled
%Control: page (0) single
%Control: year (1) truncated
%Control: production of eprint (0) enabled
\begin{thebibliography}{26}%
\makeatletter
\providecommand \@ifxundefined [1]{%
 \@ifx{#1\undefined}
}%
\providecommand \@ifnum [1]{%
 \ifnum #1\expandafter \@firstoftwo
 \else \expandafter \@secondoftwo
 \fi
}%
\providecommand \@ifx [1]{%
 \ifx #1\expandafter \@firstoftwo
 \else \expandafter \@secondoftwo
 \fi
}%
\providecommand \natexlab [1]{#1}%
\providecommand \enquote  [1]{``#1''}%
\providecommand \bibnamefont  [1]{#1}%
\providecommand \bibfnamefont [1]{#1}%
\providecommand \citenamefont [1]{#1}%
\providecommand \href@noop [0]{\@secondoftwo}%
\providecommand \href [0]{\begingroup \@sanitize@url \@href}%
\providecommand \@href[1]{\@@startlink{#1}\@@href}%
\providecommand \@@href[1]{\endgroup#1\@@endlink}%
\providecommand \@sanitize@url [0]{\catcode `\\12\catcode `\$12\catcode
  `\&12\catcode `\#12\catcode `\^12\catcode `\_12\catcode `\%12\relax}%
\providecommand \@@startlink[1]{}%
\providecommand \@@endlink[0]{}%
\providecommand \url  [0]{\begingroup\@sanitize@url \@url }%
\providecommand \@url [1]{\endgroup\@href {#1}{\urlprefix }}%
\providecommand \urlprefix  [0]{URL }%
\providecommand \Eprint [0]{\href }%
\providecommand \doibase [0]{http://dx.doi.org/}%
\providecommand \selectlanguage [0]{\@gobble}%
\providecommand \bibinfo  [0]{\@secondoftwo}%
\providecommand \bibfield  [0]{\@secondoftwo}%
\providecommand \translation [1]{[#1]}%
\providecommand \BibitemOpen [0]{}%
\providecommand \bibitemStop [0]{}%
\providecommand \bibitemNoStop [0]{.\EOS\space}%
\providecommand \EOS [0]{\spacefactor3000\relax}%
\providecommand \BibitemShut  [1]{\csname bibitem#1\endcsname}%
\let\auto@bib@innerbib\@empty
%</preamble>
\bibitem [{\citenamefont {Naidyuk}\ and\ \citenamefont {Yanson}()}]{Naidyuk}%
  \BibitemOpen
  \bibfield  {author} {\bibinfo {author} {\bibfnamefont {Y.~G.}\ \bibnamefont
  {Naidyuk}}\ and\ \bibinfo {author} {\bibfnamefont {I.~K.}\ \bibnamefont
  {Yanson}},\ }\href@noop {} {\bibinfo  {journal} {2005 Point-Contact
  Spectroscopy (New York: Springer)}\ }\BibitemShut {NoStop}%
\bibitem [{\citenamefont {Andreev}(1964)}]{Andreev}%
  \BibitemOpen
\bibfield  {journal} {  }\bibfield  {author} {\bibinfo {author} {\bibfnamefont
  {A.~F.}\ \bibnamefont {Andreev}},\ }\href@noop {} {\bibfield  {journal}
  {\bibinfo  {journal} {Sov. Phys. JETP}\ }\textbf {\bibinfo {volume} {19}},\
  \bibinfo {pages} {1228} (\bibinfo {year} {1964})}\BibitemShut {NoStop}%
\bibitem [{\citenamefont {Blonder}\ \emph {et~al.}(1982)\citenamefont
  {Blonder}, \citenamefont {Tinkham},\ and\ \citenamefont
  {Klapwijk}}]{Blonder}%
  \BibitemOpen
  \bibfield  {author} {\bibinfo {author} {\bibfnamefont {G.~E.}\ \bibnamefont
  {Blonder}}, \bibinfo {author} {\bibfnamefont {M.}~\bibnamefont {Tinkham}}, \
  and\ \bibinfo {author} {\bibfnamefont {T.~M.}\ \bibnamefont {Klapwijk}},\
  }\href@noop {} {\bibfield  {journal} {\bibinfo  {journal} {Phys. Rev. B}\
  }\textbf {\bibinfo {volume} {25}},\ \bibinfo {pages} {4515} (\bibinfo {year}
  {1982})}\BibitemShut {NoStop}%
\bibitem [{\citenamefont {Park}\ \emph {et~al.}(2008)\citenamefont {Park},
  \citenamefont {Sarrao}, \citenamefont {Thompson},\ and\ \citenamefont
  {Greene}}]{WKPark}%
  \BibitemOpen
  \bibfield  {author} {\bibinfo {author} {\bibfnamefont {W.~K.}\ \bibnamefont
  {Park}}, \bibinfo {author} {\bibfnamefont {J.~L.}\ \bibnamefont {Sarrao}},
  \bibinfo {author} {\bibfnamefont {J.~D.}\ \bibnamefont {Thompson}}, \ and\
  \bibinfo {author} {\bibfnamefont {L.~H.}\ \bibnamefont {Greene}},\
  }\href@noop {} {\bibfield  {journal} {\bibinfo  {journal} {Phys. Rev. Lett.}\
  }\textbf {\bibinfo {volume} {100}},\ \bibinfo {pages} {177001} (\bibinfo
  {year} {2008})}\BibitemShut {NoStop}%
\bibitem [{\citenamefont {Gonnelli}\ \emph {et~al.}(2002)\citenamefont
  {Gonnelli}, \citenamefont {Daghero}, \citenamefont {Ummarino}, \citenamefont
  {Stepanov}, \citenamefont {Jun}, \citenamefont {Kazakov},\ and\ \citenamefont
  {Karpinski}}]{MgB2}%
  \BibitemOpen
  \bibfield  {author} {\bibinfo {author} {\bibfnamefont {R.~S.}\ \bibnamefont
  {Gonnelli}}, \bibinfo {author} {\bibfnamefont {D.}~\bibnamefont {Daghero}},
  \bibinfo {author} {\bibfnamefont {G.~A.}\ \bibnamefont {Ummarino}}, \bibinfo
  {author} {\bibfnamefont {V.~A.}\ \bibnamefont {Stepanov}}, \bibinfo {author}
  {\bibfnamefont {J.}~\bibnamefont {Jun}}, \bibinfo {author} {\bibfnamefont
  {S.~M.}\ \bibnamefont {Kazakov}}, \ and\ \bibinfo {author} {\bibfnamefont
  {J.}~\bibnamefont {Karpinski}},\ }\href@noop {} {\bibfield  {journal}
  {\bibinfo  {journal} {Phys. Rev. Lett.}\ }\textbf {\bibinfo {volume} {89}},\
  \bibinfo {pages} {247004} (\bibinfo {year} {2002})}\BibitemShut {NoStop}%
\bibitem [{\citenamefont {Daghero}\ \emph {et~al.}(2011)\citenamefont
  {Daghero}, \citenamefont {Tortello}, \citenamefont {Ummarino},\ and\
  \citenamefont {Gonnelli}}]{ROPP}%
  \BibitemOpen
  \bibfield  {author} {\bibinfo {author} {\bibfnamefont {D.}~\bibnamefont
  {Daghero}}, \bibinfo {author} {\bibfnamefont {M.}~\bibnamefont {Tortello}},
  \bibinfo {author} {\bibfnamefont {G.~A.}\ \bibnamefont {Ummarino}}, \ and\
  \bibinfo {author} {\bibfnamefont {R.~S.}\ \bibnamefont {Gonnelli}},\
  }\href@noop {} {\bibfield  {journal} {\bibinfo  {journal} {Rep. Prog. Phys.}\
  }\textbf {\bibinfo {volume} {74}},\ \bibinfo {pages} {124509} (\bibinfo
  {year} {2011})}\BibitemShut {NoStop}%
\bibitem [{\citenamefont {Samuely}\ \emph {et~al.}(2009)\citenamefont
  {Samuely}, \citenamefont {Pribulova}, \citenamefont {Szabo}, \citenamefont
  {Pristas}, \citenamefont {Bud'ko},\ and\ \citenamefont {Canfield}}]{Samuely}%
  \BibitemOpen
  \bibfield  {author} {\bibinfo {author} {\bibfnamefont {P.}~\bibnamefont
  {Samuely}}, \bibinfo {author} {\bibfnamefont {Z.}~\bibnamefont {Pribulova}},
  \bibinfo {author} {\bibfnamefont {P.}~\bibnamefont {Szabo}}, \bibinfo
  {author} {\bibfnamefont {G.}~\bibnamefont {Pristas}}, \bibinfo {author}
  {\bibfnamefont {S.~L.}\ \bibnamefont {Bud'ko}}, \ and\ \bibinfo {author}
  {\bibfnamefont {P.~C.}\ \bibnamefont {Canfield}},\ }\href@noop {} {\bibfield
  {journal} {\bibinfo  {journal} {Physica C}\ }\textbf {\bibinfo {volume}
  {469}},\ \bibinfo {pages} {507} (\bibinfo {year} {2009})}\BibitemShut
  {NoStop}%
\bibitem [{\citenamefont {Tortello}\ \emph {et~al.}(2010)\citenamefont
  {Tortello}, \citenamefont {Daghero}, \citenamefont {Ummarino}, \citenamefont
  {Stepanov}, \citenamefont {Jiang}, \citenamefont {Weiss}, \citenamefont
  {Hellstrom},\ and\ \citenamefont {Gonnelli}}]{Tortello}%
  \BibitemOpen
  \bibfield  {author} {\bibinfo {author} {\bibfnamefont {M.}~\bibnamefont
  {Tortello}}, \bibinfo {author} {\bibfnamefont {D.}~\bibnamefont {Daghero}},
  \bibinfo {author} {\bibfnamefont {G.~A.}\ \bibnamefont {Ummarino}}, \bibinfo
  {author} {\bibfnamefont {V.~A.}\ \bibnamefont {Stepanov}}, \bibinfo {author}
  {\bibfnamefont {J.}~\bibnamefont {Jiang}}, \bibinfo {author} {\bibfnamefont
  {J.~D.}\ \bibnamefont {Weiss}}, \bibinfo {author} {\bibfnamefont {E.~E.}\
  \bibnamefont {Hellstrom}}, \ and\ \bibinfo {author} {\bibfnamefont {R.~S.}\
  \bibnamefont {Gonnelli}},\ }\href@noop {} {\bibfield  {journal} {\bibinfo
  {journal} {Phys. Rev. Lett.}\ }\textbf {\bibinfo {volume} {105}},\ \bibinfo
  {pages} {237002} (\bibinfo {year} {2010})}\BibitemShut {NoStop}%
\bibitem [{\citenamefont {Johnston}(2010)}]{Johnston}%
  \BibitemOpen
  \bibfield  {author} {\bibinfo {author} {\bibfnamefont {D.~C.}\ \bibnamefont
  {Johnston}},\ }\href@noop {} {\bibfield  {journal} {\bibinfo  {journal}
  {Advances in Physics}\ }\textbf {\bibinfo {volume} {59}},\ \bibinfo {pages}
  {803} (\bibinfo {year} {2010})}\BibitemShut {NoStop}%
\bibitem [{\citenamefont {Fisher}\ \emph {et~al.}(2011)\citenamefont {Fisher},
  \citenamefont {Degiorgi},\ and\ \citenamefont {Shen}}]{Fisher}%
  \BibitemOpen
  \bibfield  {author} {\bibinfo {author} {\bibfnamefont {I.~R.}\ \bibnamefont
  {Fisher}}, \bibinfo {author} {\bibfnamefont {L.}~\bibnamefont {Degiorgi}}, \
  and\ \bibinfo {author} {\bibfnamefont {Z.~X.}\ \bibnamefont {Shen}},\
  }\href@noop {} {\bibfield  {journal} {\bibinfo  {journal} {Rep. Prog. Phys.}\
  }\textbf {\bibinfo {volume} {74}},\ \bibinfo {pages} {124506} (\bibinfo
  {year} {2011})}\BibitemShut {NoStop}%
\bibitem [{\citenamefont {Brinkman}\ \emph {et~al.}(2002)\citenamefont
  {Brinkman}, \citenamefont {Golubov}, \citenamefont {Rogalla}, \citenamefont
  {Dolgov}, \citenamefont {Kortus}, \citenamefont {Kong}, \citenamefont
  {Jepsen},\ and\ \citenamefont {Andersen}}]{Brinkman}%
  \BibitemOpen
  \bibfield  {author} {\bibinfo {author} {\bibfnamefont {A.}~\bibnamefont
  {Brinkman}}, \bibinfo {author} {\bibfnamefont {A.~A.}\ \bibnamefont
  {Golubov}}, \bibinfo {author} {\bibfnamefont {H.}~\bibnamefont {Rogalla}},
  \bibinfo {author} {\bibfnamefont {O.~V.}\ \bibnamefont {Dolgov}}, \bibinfo
  {author} {\bibfnamefont {J.}~\bibnamefont {Kortus}}, \bibinfo {author}
  {\bibfnamefont {Y.}~\bibnamefont {Kong}}, \bibinfo {author} {\bibfnamefont
  {O.}~\bibnamefont {Jepsen}}, \ and\ \bibinfo {author} {\bibfnamefont {O.~K.}\
  \bibnamefont {Andersen}},\ }\href@noop {} {\bibfield  {journal} {\bibinfo
  {journal} {Phys. Rev. B}\ }\textbf {\bibinfo {volume} {65}},\ \bibinfo
  {pages} {180517(R)} (\bibinfo {year} {2002})}\BibitemShut {NoStop}%
\bibitem [{\citenamefont {Sebastian}\ \emph {et~al.}(2008)\citenamefont
  {Sebastian}, \citenamefont {Gillett}, \citenamefont {Harrison}, \citenamefont
  {Lau}, \citenamefont {Singh}, \citenamefont {Mielke},\ and\ \citenamefont
  {Lonzarich}}]{Sebastian}%
  \BibitemOpen
  \bibfield  {author} {\bibinfo {author} {\bibfnamefont {S.~E.}\ \bibnamefont
  {Sebastian}}, \bibinfo {author} {\bibfnamefont {J.}~\bibnamefont {Gillett}},
  \bibinfo {author} {\bibfnamefont {N.}~\bibnamefont {Harrison}}, \bibinfo
  {author} {\bibfnamefont {P.~H.~C.}\ \bibnamefont {Lau}}, \bibinfo {author}
  {\bibfnamefont {D.~J.}\ \bibnamefont {Singh}}, \bibinfo {author}
  {\bibfnamefont {C.~H.}\ \bibnamefont {Mielke}}, \ and\ \bibinfo {author}
  {\bibfnamefont {G.~G.}\ \bibnamefont {Lonzarich}},\ }\href@noop {} {\bibfield
   {journal} {\bibinfo  {journal} {J. Phys.: Cond. Matt.}\ }\textbf {\bibinfo
  {volume} {20}},\ \bibinfo {pages} {422203} (\bibinfo {year}
  {2008})}\BibitemShut {NoStop}%
\bibitem [{\citenamefont {Gillett}\ \emph {et~al.}()\citenamefont {Gillett},
  \citenamefont {Das}, \citenamefont {Syers}, \citenamefont {Ming},
  \citenamefont {Espeso}, \citenamefont {Petrone},\ and\ \citenamefont
  {Sebastian}}]{Gillett}%
  \BibitemOpen
  \bibfield  {author} {\bibinfo {author} {\bibfnamefont {J.}~\bibnamefont
  {Gillett}}, \bibinfo {author} {\bibfnamefont {S.~D.}\ \bibnamefont {Das}},
  \bibinfo {author} {\bibfnamefont {P.}~\bibnamefont {Syers}}, \bibinfo
  {author} {\bibfnamefont {A.~K.~T.}\ \bibnamefont {Ming}}, \bibinfo {author}
  {\bibfnamefont {J.~I.}\ \bibnamefont {Espeso}}, \bibinfo {author}
  {\bibfnamefont {C.~M.}\ \bibnamefont {Petrone}}, \ and\ \bibinfo {author}
  {\bibfnamefont {S.~E.}\ \bibnamefont {Sebastian}},\ }\href@noop {} {\bibinfo
  {journal} {arXiv:1005.1330v1}\ }\BibitemShut {NoStop}%
\bibitem [{\citenamefont {Chung}\ \emph {et~al.}()\citenamefont {Chung} \emph
  {et~al.}}]{Duck}%
  \BibitemOpen
\bibfield  {journal} {  }\bibfield  {author} {\bibinfo {author} {\bibfnamefont
  {D.~Y.}\ \bibnamefont {Chung}} \emph {et~al.},\ }\href@noop {} {\bibinfo
  {journal} {unpublished}\ }\BibitemShut {NoStop}%
\bibitem [{\citenamefont {Arham}\ \emph {et~al.}(2012)\citenamefont {Arham},
  \citenamefont {Hunt}, \citenamefont {Park}, \citenamefont {Gillett},
  \citenamefont {Das}, \citenamefont {Sebastian}, \citenamefont {Xu},
  \citenamefont {Wen}, \citenamefont {Lin}, \citenamefont {Li}, \citenamefont
  {Gu}, \citenamefont {Thaler}, \citenamefont {Ran}, \citenamefont {Bud'ko},
  \citenamefont {Canfield}, \citenamefont {Chung}, \citenamefont {Kanatzidis},\
  and\ \citenamefont {Greene}}]{Arham}%
  \BibitemOpen
\bibfield  {journal} {  }\bibfield  {author} {\bibinfo {author} {\bibfnamefont
  {H.~Z.}\ \bibnamefont {Arham}}, \bibinfo {author} {\bibfnamefont {C.~R.}\
  \bibnamefont {Hunt}}, \bibinfo {author} {\bibfnamefont {W.~K.}\ \bibnamefont
  {Park}}, \bibinfo {author} {\bibfnamefont {J.}~\bibnamefont {Gillett}},
  \bibinfo {author} {\bibfnamefont {S.~D.}\ \bibnamefont {Das}}, \bibinfo
  {author} {\bibfnamefont {S.~E.}\ \bibnamefont {Sebastian}}, \bibinfo {author}
  {\bibfnamefont {Z.~J.}\ \bibnamefont {Xu}}, \bibinfo {author} {\bibfnamefont
  {J.~S.}\ \bibnamefont {Wen}}, \bibinfo {author} {\bibfnamefont {Z.~W.}\
  \bibnamefont {Lin}}, \bibinfo {author} {\bibfnamefont {Q.}~\bibnamefont
  {Li}}, \bibinfo {author} {\bibfnamefont {G.}~\bibnamefont {Gu}}, \bibinfo
  {author} {\bibfnamefont {A.}~\bibnamefont {Thaler}}, \bibinfo {author}
  {\bibfnamefont {S.}~\bibnamefont {Ran}}, \bibinfo {author} {\bibfnamefont
  {S.~L.}\ \bibnamefont {Bud'ko}}, \bibinfo {author} {\bibfnamefont {P.~C.}\
  \bibnamefont {Canfield}}, \bibinfo {author} {\bibfnamefont {D.~Y.}\
  \bibnamefont {Chung}}, \bibinfo {author} {\bibfnamefont {M.~G.}\ \bibnamefont
  {Kanatzidis}}, \ and\ \bibinfo {author} {\bibfnamefont {L.~H.}\ \bibnamefont
  {Greene}},\ }\href@noop {} {\bibfield  {journal} {\bibinfo  {journal} {Phys.
  Rev. B}\ }\textbf {\bibinfo {volume} {85}},\ \bibinfo {pages} {214515}
  (\bibinfo {year} {2012})}\BibitemShut {NoStop}%
\bibitem [{\citenamefont {Dynes}\ \emph {et~al.}(1978)\citenamefont {Dynes},
  \citenamefont {Narayanamurti},\ and\ \citenamefont {Garno}}]{Dynes}%
  \BibitemOpen
  \bibfield  {author} {\bibinfo {author} {\bibfnamefont {R.~C.}\ \bibnamefont
  {Dynes}}, \bibinfo {author} {\bibfnamefont {V.}~\bibnamefont
  {Narayanamurti}}, \ and\ \bibinfo {author} {\bibfnamefont {J.~P.}\
  \bibnamefont {Garno}},\ }\href@noop {} {\bibfield  {journal} {\bibinfo
  {journal} {Phys. Rev. Lett.}\ }\textbf {\bibinfo {volume} {41}},\ \bibinfo
  {pages} {1509} (\bibinfo {year} {1978})}\BibitemShut {NoStop}%
\bibitem [{\citenamefont {Teague}\ \emph {et~al.}(2011)\citenamefont {Teague},
  \citenamefont {Drayna}, \citenamefont {Lockhart}, \citenamefont {Cheng},
  \citenamefont {Shen}, \citenamefont {Wen},\ and\ \citenamefont
  {Yeh}}]{Teague}%
  \BibitemOpen
  \bibfield  {author} {\bibinfo {author} {\bibfnamefont {M.~L.}\ \bibnamefont
  {Teague}}, \bibinfo {author} {\bibfnamefont {G.~K.}\ \bibnamefont {Drayna}},
  \bibinfo {author} {\bibfnamefont {G.~P.}\ \bibnamefont {Lockhart}}, \bibinfo
  {author} {\bibfnamefont {P.}~\bibnamefont {Cheng}}, \bibinfo {author}
  {\bibfnamefont {B.}~\bibnamefont {Shen}}, \bibinfo {author} {\bibfnamefont
  {H.-H.}\ \bibnamefont {Wen}}, \ and\ \bibinfo {author} {\bibfnamefont
  {N.-C.}\ \bibnamefont {Yeh}},\ }\href@noop {} {\bibfield  {journal} {\bibinfo
   {journal} {Phys. Rev. Lett.}\ }\textbf {\bibinfo {volume} {106}},\ \bibinfo
  {pages} {087004} (\bibinfo {year} {2011})}\BibitemShut {NoStop}%
\bibitem [{\citenamefont {Terashima}\ \emph {et~al.}(2009)\citenamefont
  {Terashima}, \citenamefont {Sekiba}, \citenamefont {Bowen}, \citenamefont
  {Nakayama}, \citenamefont {Kawahara}, \citenamefont {Sato}, \citenamefont
  {Richard}, \citenamefont {Xu}, \citenamefont {Li}, \citenamefont {Cao},
  \citenamefont {Xu}, \citenamefont {Ding},\ and\ \citenamefont
  {Takahashi}}]{Terashima}%
  \BibitemOpen
  \bibfield  {author} {\bibinfo {author} {\bibfnamefont {K.}~\bibnamefont
  {Terashima}}, \bibinfo {author} {\bibfnamefont {Y.}~\bibnamefont {Sekiba}},
  \bibinfo {author} {\bibfnamefont {J.~H.}\ \bibnamefont {Bowen}}, \bibinfo
  {author} {\bibfnamefont {K.}~\bibnamefont {Nakayama}}, \bibinfo {author}
  {\bibfnamefont {T.}~\bibnamefont {Kawahara}}, \bibinfo {author}
  {\bibfnamefont {T.}~\bibnamefont {Sato}}, \bibinfo {author} {\bibfnamefont
  {P.}~\bibnamefont {Richard}}, \bibinfo {author} {\bibfnamefont {Y.-M.}\
  \bibnamefont {Xu}}, \bibinfo {author} {\bibfnamefont {L.~J.}\ \bibnamefont
  {Li}}, \bibinfo {author} {\bibfnamefont {G.~H.}\ \bibnamefont {Cao}},
  \bibinfo {author} {\bibfnamefont {Z.-A.}\ \bibnamefont {Xu}}, \bibinfo
  {author} {\bibfnamefont {H.}~\bibnamefont {Ding}}, \ and\ \bibinfo {author}
  {\bibfnamefont {T.}~\bibnamefont {Takahashi}},\ }\href@noop {} {\bibfield
  {journal} {\bibinfo  {journal} {PNAS}\ }\textbf {\bibinfo {volume} {106}},\
  \bibinfo {pages} {7330} (\bibinfo {year} {2009})}\BibitemShut {NoStop}%
\bibitem [{\citenamefont {Basov}\ and\ \citenamefont
  {Chubukov}(2011)}]{Chubukov}%
  \BibitemOpen
  \bibfield  {author} {\bibinfo {author} {\bibfnamefont {D.~N.}\ \bibnamefont
  {Basov}}\ and\ \bibinfo {author} {\bibfnamefont {A.~V.}\ \bibnamefont
  {Chubukov}},\ }\href@noop {} {\bibfield  {journal} {\bibinfo  {journal}
  {Nature Physics}\ }\textbf {\bibinfo {volume} {7}},\ \bibinfo {pages} {272}
  (\bibinfo {year} {2011})}\BibitemShut {NoStop}%
\bibitem [{\citenamefont {Ding}\ \emph {et~al.}(2008)\citenamefont {Ding},
  \citenamefont {Richard}, \citenamefont {Nakayama}, \citenamefont {Sugawara},
  \citenamefont {Arakane}, \citenamefont {Sekiba}, \citenamefont {Takayama},
  \citenamefont {Souma}, \citenamefont {Sato}, \citenamefont {Takahashi},
  \citenamefont {Wang}, \citenamefont {Dai}, \citenamefont {Fang},
  \citenamefont {Chen}, \citenamefont {Luo},\ and\ \citenamefont
  {Wang}}]{Ding}%
  \BibitemOpen
  \bibfield  {author} {\bibinfo {author} {\bibfnamefont {H.}~\bibnamefont
  {Ding}}, \bibinfo {author} {\bibfnamefont {P.}~\bibnamefont {Richard}},
  \bibinfo {author} {\bibfnamefont {K.}~\bibnamefont {Nakayama}}, \bibinfo
  {author} {\bibfnamefont {K.}~\bibnamefont {Sugawara}}, \bibinfo {author}
  {\bibfnamefont {T.}~\bibnamefont {Arakane}}, \bibinfo {author} {\bibfnamefont
  {Y.}~\bibnamefont {Sekiba}}, \bibinfo {author} {\bibfnamefont
  {A.}~\bibnamefont {Takayama}}, \bibinfo {author} {\bibfnamefont
  {S.}~\bibnamefont {Souma}}, \bibinfo {author} {\bibfnamefont
  {T.}~\bibnamefont {Sato}}, \bibinfo {author} {\bibfnamefont {T.}~\bibnamefont
  {Takahashi}}, \bibinfo {author} {\bibfnamefont {Z.}~\bibnamefont {Wang}},
  \bibinfo {author} {\bibfnamefont {X.}~\bibnamefont {Dai}}, \bibinfo {author}
  {\bibfnamefont {Z.}~\bibnamefont {Fang}}, \bibinfo {author} {\bibfnamefont
  {G.~F.}\ \bibnamefont {Chen}}, \bibinfo {author} {\bibfnamefont {J.~L.}\
  \bibnamefont {Luo}}, \ and\ \bibinfo {author} {\bibfnamefont {N.~L.}\
  \bibnamefont {Wang}},\ }\href@noop {} {\bibfield  {journal} {\bibinfo
  {journal} {Europhys. Lett.}\ }\textbf {\bibinfo {volume} {83}},\ \bibinfo
  {pages} {47001} (\bibinfo {year} {2008})}\BibitemShut {NoStop}%
\bibitem [{\citenamefont {Golubov}\ \emph {et~al.}(2009)\citenamefont
  {Golubov}, \citenamefont {Brinkman}, \citenamefont {Tanaka}, \citenamefont
  {Mazin},\ and\ \citenamefont {Dolgov}}]{Golubov}%
  \BibitemOpen
  \bibfield  {author} {\bibinfo {author} {\bibfnamefont {A.~A.}\ \bibnamefont
  {Golubov}}, \bibinfo {author} {\bibfnamefont {A.}~\bibnamefont {Brinkman}},
  \bibinfo {author} {\bibfnamefont {Y.}~\bibnamefont {Tanaka}}, \bibinfo
  {author} {\bibfnamefont {I.~I.}\ \bibnamefont {Mazin}}, \ and\ \bibinfo
  {author} {\bibfnamefont {O.~V.}\ \bibnamefont {Dolgov}},\ }\href@noop {}
  {\bibfield  {journal} {\bibinfo  {journal} {Phys. Rev. Lett.}\ }\textbf
  {\bibinfo {volume} {103}},\ \bibinfo {pages} {077003} (\bibinfo {year}
  {2009})}\BibitemShut {NoStop}%
\bibitem [{\citenamefont {Sperstad}\ \emph {et~al.}(2009)\citenamefont
  {Sperstad}, \citenamefont {Linder},\ and\ \citenamefont {Sudbo}}]{Sudbo}%
  \BibitemOpen
  \bibfield  {author} {\bibinfo {author} {\bibfnamefont {I.~B.}\ \bibnamefont
  {Sperstad}}, \bibinfo {author} {\bibfnamefont {J.}~\bibnamefont {Linder}}, \
  and\ \bibinfo {author} {\bibfnamefont {A.}~\bibnamefont {Sudbo}},\
  }\href@noop {} {\bibfield  {journal} {\bibinfo  {journal} {Phys. Rev. B}\
  }\textbf {\bibinfo {volume} {80}},\ \bibinfo {pages} {144507} (\bibinfo
  {year} {2009})}\BibitemShut {NoStop}%
\bibitem [{\citenamefont {Arham}\ \emph {et~al.}()\citenamefont {Arham} \emph
  {et~al.}}]{Supplement}%
  \BibitemOpen
  \bibfield  {author} {\bibinfo {author} {\bibfnamefont {H.~Z.}\ \bibnamefont
  {Arham}} \emph {et~al.},\ }\href@noop {} {\bibinfo  {journal} {See
  Supplemental Material at [URL will be inserted by publisher] for discussion
  on Excess Current Calculation and Inhomogeneous Doping Model}\ }\BibitemShut
  {NoStop}%
\bibitem [{\citenamefont {Park}\ \emph {et~al.}()\citenamefont {Park},
  \citenamefont {Hunt}, \citenamefont {Arham}, \citenamefont {Xu},
  \citenamefont {Wen}, \citenamefont {Lin}, \citenamefont {Li}, \citenamefont
  {Gu},\ and\ \citenamefont {Greene}}]{FeTeSe}%
  \BibitemOpen
\bibfield  {journal} {  }\bibfield  {author} {\bibinfo {author} {\bibfnamefont
  {W.~K.}\ \bibnamefont {Park}}, \bibinfo {author} {\bibfnamefont {C.~R.}\
  \bibnamefont {Hunt}}, \bibinfo {author} {\bibfnamefont {H.~Z.}\ \bibnamefont
  {Arham}}, \bibinfo {author} {\bibfnamefont {Z.~J.}\ \bibnamefont {Xu}},
  \bibinfo {author} {\bibfnamefont {J.~S.}\ \bibnamefont {Wen}}, \bibinfo
  {author} {\bibfnamefont {Z.~W.}\ \bibnamefont {Lin}}, \bibinfo {author}
  {\bibfnamefont {Q.}~\bibnamefont {Li}}, \bibinfo {author} {\bibfnamefont
  {G.~D.}\ \bibnamefont {Gu}}, \ and\ \bibinfo {author} {\bibfnamefont {L.~H.}\
  \bibnamefont {Greene}},\ }\href@noop {} {\bibinfo  {journal}
  {arXiv:1005.0190}\ }\BibitemShut {NoStop}%
\bibitem [{\citenamefont {Sheet}\ \emph {et~al.}(2010)\citenamefont {Sheet},
  \citenamefont {Mehta}, \citenamefont {Dikin}, \citenamefont {Lee},
  \citenamefont {Bark}, \citenamefont {Jiang}, \citenamefont {Weiss},
  \citenamefont {Hellstrom}, \citenamefont {Rzchowski}, \citenamefont {Eom},\
  and\ \citenamefont {Chandrasekhar}}]{Goutam}%
  \BibitemOpen
\bibfield  {journal} {  }\bibfield  {author} {\bibinfo {author} {\bibfnamefont
  {G.}~\bibnamefont {Sheet}}, \bibinfo {author} {\bibfnamefont
  {M.}~\bibnamefont {Mehta}}, \bibinfo {author} {\bibfnamefont {D.~A.}\
  \bibnamefont {Dikin}}, \bibinfo {author} {\bibfnamefont {S.}~\bibnamefont
  {Lee}}, \bibinfo {author} {\bibfnamefont {C.~W.}\ \bibnamefont {Bark}},
  \bibinfo {author} {\bibfnamefont {J.}~\bibnamefont {Jiang}}, \bibinfo
  {author} {\bibfnamefont {J.~D.}\ \bibnamefont {Weiss}}, \bibinfo {author}
  {\bibfnamefont {E.~E.}\ \bibnamefont {Hellstrom}}, \bibinfo {author}
  {\bibfnamefont {M.~S.}\ \bibnamefont {Rzchowski}}, \bibinfo {author}
  {\bibfnamefont {C.~B.}\ \bibnamefont {Eom}}, \ and\ \bibinfo {author}
  {\bibfnamefont {V.}~\bibnamefont {Chandrasekhar}},\ }\href@noop {} {\bibfield
   {journal} {\bibinfo  {journal} {Phys. Rev. Lett.}\ }\textbf {\bibinfo
  {volume} {105}},\ \bibinfo {pages} {167003} (\bibinfo {year}
  {2010})}\BibitemShut {NoStop}%
\bibitem [{\citenamefont {Lee}\ and\ \citenamefont
  {Phillips}(2012)}]{Weicheng}%
  \BibitemOpen
  \bibfield  {author} {\bibinfo {author} {\bibfnamefont {W.-C.}\ \bibnamefont
  {Lee}}\ and\ \bibinfo {author} {\bibfnamefont {P.~W.}\ \bibnamefont
  {Phillips}},\ }\href@noop {} {\bibfield  {journal} {\bibinfo  {journal}
  {Phys. Rev. B}\ }\textbf {\bibinfo {volume} {86}},\ \bibinfo {pages} {245113}
  (\bibinfo {year} {2012})}\BibitemShut {NoStop}%
\end{thebibliography}%
\end{document}

% Use the \preprint command to place your local institutional report
% number in the upper righthand corner of the title page in preprint mode.
% Multiple \preprint commands are allowed.
% Use the 'preprintnumbers' class option to override journal defaults
% to display numbers if necessary
%\preprint{}

%Title of paper
\title{Supplemental Material for\\ Andreev Reflection Like Enhancement Above Bulk $T_c$ in Electron Underdoped Iron Arsenides}

\author{H. Z. Arham$^1$}
\email[E-mail:]{arham1@illinois.edu}
\author{C. R. Hunt,$^1$ J. Gillett,$^2$ S. D. Das,$^2$ S. E. Sebastian$^2$}
\author{D. Y. Chung,$^3$ M. G. Kanatzidis$^3$}
\author{L. H. Greene$^1$}

\affiliation{\\\textsuperscript{$1$}Department of Physics and the Frederick Seitz Material Research Laboratory, University of Illinois at Urbana-Champaign, Urbana, Illinois 61801, USA
\\\textsuperscript{$2$}Cavendish Laboratory, J. J. Thomson Ave, University of Cambridge, UK
\\\textsuperscript{$3$}Materials Science Division, Argonne National Laboratory, Argonne, IL 60439, USA}

\date{\today}
\pacs{}
\maketitle

\section{Excess Current Calculation}
Andreev reflection causes an increase in the current transmitted across a normal metal-superconductor point contact junction. $I_{exc}$ is defined as the extra current that flows through the junction when compared with its non-superconducting state. To calculate this current we use the equation:

\begin{equation}
I_{exc}(T)=\int^{V>>\Delta}_{-V<<-\Delta} \frac{dI}{dV}(T)\mathrm{d}V - \int^{V>>\Delta}_{-V<<-\Delta} \frac{dI}{dV}(T \geq T_c^{onset})\mathrm{d}V
\end{equation}
 
\begin{figure*}[thbp]
		\includegraphics[scale=0.7]{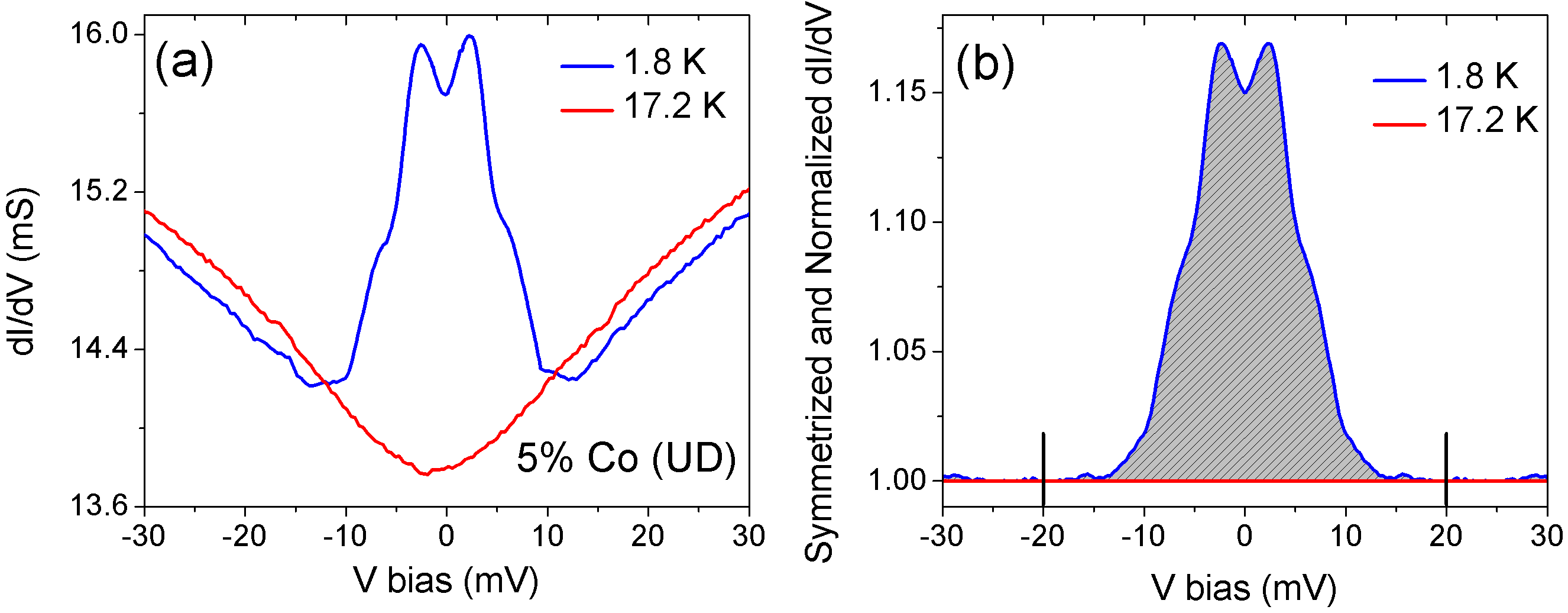}
	\caption{(a) Raw $dI/dV$ curves taken on $\rm{Ba(Fe_{0.95}Co_{0.05})_2As_2}$ at 1.8 K (blue) and 17.2 K (red). (b) The same curves, after they have been symmetrized and normalized to $dI/dV$ at 17.2 K. $I_{exc}$ is represented by the shaded gray area on the plot. It is calculated by integrating the area under the curves and subtracting the value at 17.2 K from the value at 1.8 K. The integration limits are set at $\pm$ 20 mV, represented by the short vertical black lines. At biases higher than 20 mV the two $dI/dV$ curves are nearly identical.}
	\label{fig:100}
\end{figure*}  

We illustrate how this integration is performed for $\rm{Ba(Fe_{0.95}Co_{0.05})_2As_2}$ in Figure 1.

Figure 1a shows the raw $dI/dV$ curves taken at 1.8 K (blue, lowest temperature attained for this junction) and 17.2 K (red, $T$ $\sim$ $T_c^{onset}$). Figure 1b shows the same curves after they has been symmetrized and normalized with the curve at 17.2 K. Symmetrization is carried out by taking the average of the $dI/dV$ values at positive and negative biases. 

The next step is to integrate the area under the curves and subtract the current at 17 K from the current at 1.8 K. We choose the integration limit to be from -20mV to +20mV since at biases higher than that Andreev reflection dies out and the two $dI/dV$ curves are nearly identical. The gray shaded area in Figure 1b represents the final $I_{exc}$ that we obtain. 

This same procedure is repeated for all our crystals at various temperatures. Figure 4a in the main text of our paper is obtained by combining all of the $I_{exc}$ data points. 

\section{Inhomogeneous Doping Model}

\begin{figure*}[thbp]
		\includegraphics[scale=0.7]{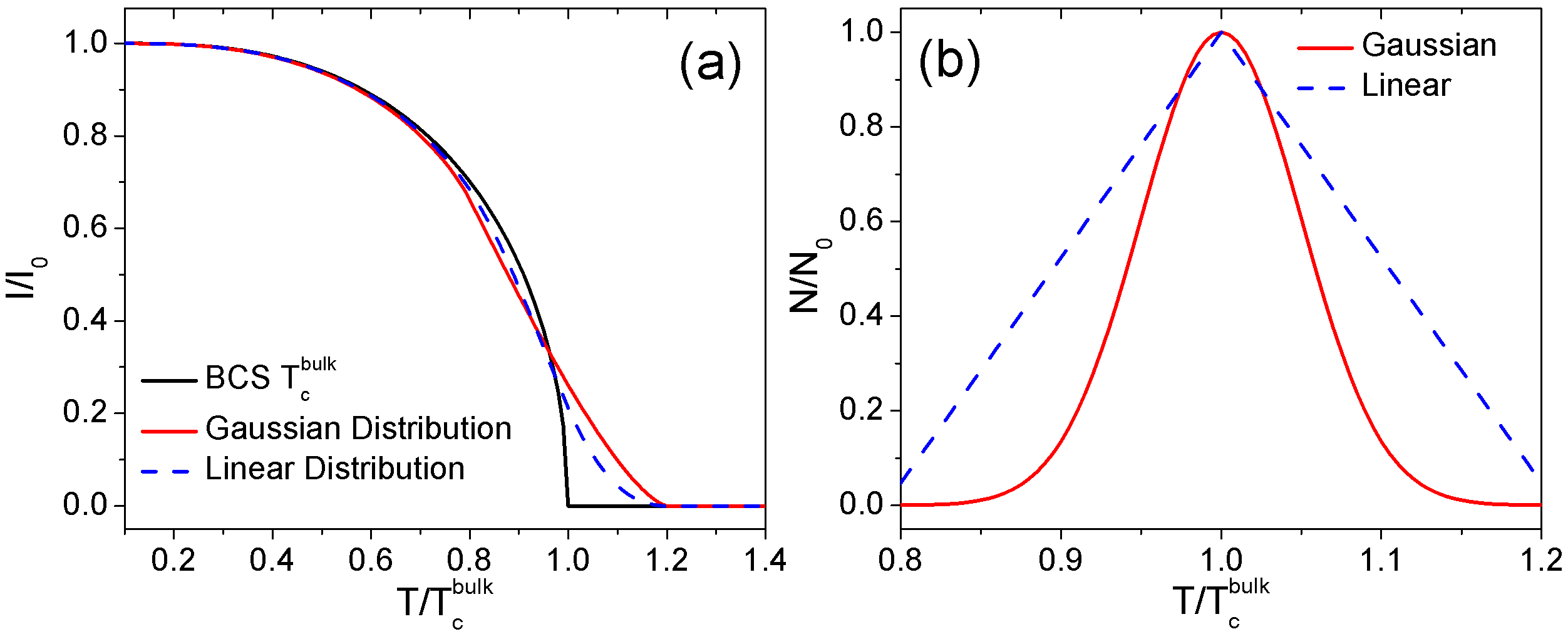}
	\caption{(a) Comparing the $I_{exc}$ calculated from Gaussian and linear distribution functions with the BCS $I_{exc}$. The distribution function $I_{exc}$ develops a tail above $T_c^{bulk}$. (b) The number of channels with a given $T_c$ for the Gaussian ($\sigma=5\%$ $T_c^{bulk}$) and the Linear distribution functions.}
	\label{fig:100}
\end{figure*}  

Our basic assumption is that our point contact junction is comprised of multiple channels and there is slight variation in the local $T_c$ values of these channels. Most of the channels start showing Andreev reflection at $T_c^{bulk}$ but some of them start Andreev reflecting above it while other start below it. We define a variable $T_c^{local}$ and set its limits to be 0.8-1.2 $T_c^{bulk}$. 

We have tried various distribution functions for $T_c^{local}$. Figure 2b shows the number of channels with a given $T_c$ for a Gaussian ($\sigma=5\%$ $T_c^{bulk}$) and a Linear distribution function. The largest number of channels superconduct at $T_c^{bulk}$ and as $T_c^{local}$ deviates more and more from $T_c^{bulk}$, the number of channels with that $T_c$ falls. 

$I_{exc}$ is calculated by summing up the excess current due to all the Andreev reflection channels. Figure 4b in the main text of our paper uses the Gaussian distribution function to calculate $I_{exc}$. In Figure 2a, we compare $I_{exc}$ from the Gaussian and linear distribution functions with BCS $I_{exc}$. The general feature is that the $I_{exc}$ for the distribution functions develops a tail above $T_c^{bulk}$. 

% Create the reference section using BibTeX:
\bibliography{myrefs}